\documentclass[preprints,article,accept,moreauthors,pdftex]{Definitions/mdpi} 
\firstpage{1} 
\pubvolume{1}
\issuenum{1}
\articlenumber{0}
\pubyear{2023}
\copyrightyear{2023}
\datereceived{} 
\dateaccepted{} 
\datepublished{} 
\hreflink{https://doi.org/} 

\usepackage{bm}

\newcommand{\aap}{Astron.\ Astrophys.}
\newcommand{\mnras}{Mon.\ Not.\ R.\ Astron.\ Soc.}

\newcommand{\prc}{Phys.\ Rev.\ \rm C}
\newcommand{\prd}{Phys.\ Rev.\ \rm D}
\newcommand{\prl}{Phys.\ Rev.\ Lett.}
\newcommand{\apj}{Astrophys.\ J.}

\newcommand{\apjs}{Astrophys.\ J.\ Suppl.\ Ser.}

\newcommand{\nphysa}{Nucl.\ Phys.\ \rm A}

\Title{Elasticity of neutron star mantle: improved compressible liquid drop model for cylindrical phases}

\TitleCitation{Elasticity of neutron star mantle: improved CLDM for cylindrical phases}

\Author{Nikita A. Zemlyakov, Andrey I. Chugunov\orcidC{}}

\AuthorNames{Nikita A. Zemlyakov, Andrey I. Chugunov}

\AuthorCitation{Zemlyakov, N.A.; Chugunov, A.I.}

\address{%
Ioffe Institute, 26 Politekhnicheskaya st., St. Petersburg 194021, Russia}

\abstract{Neutron stars are the densest objects in the Universe. They have microscopically homogeneous core and heterogeneous crust. 
In particular, there may be a specific layer inside neutron stars, the mantle, which consists of substantially non-spherical nuclei immersed in a background of relativistic degenerate electrons and quasi-free neutrons.
In this paper we reconsider transverse shear modulus for cylindrical phases of the mantle within the framework of compressible liquid drop model. 
We demonstrate that transverse shear affects the shape of nuclear clusters: their cross-section becomes elliptical. This effect reduces respective elastic constant.
Using a simple model we perform all derivations analytically and obtain the expression for the transverse shear modulus, which can be useful for astrophysical applications.}


\begin{document}


\section{Introduction}
The mantle is a specific microscopically inhomogeneous layer of neutron star, where nuclear clusters's shapes are essentially non-spherical. This layer was predicted for the first time in \cite{Ravenhall_pasta,Hashimoto_ea84} by considering a set of symmetrical pasta-like nuclear shapes -- cylinders, plates, and their inverse configurations -- along with spherical nuclei and spherical holes.
This set has become canonical and was considered in a large number of subsequent papers (e.g.,  \cite{Lorenz+93, Oyamatsu93, DHM00,Iida_ea01, Nakazato_ea11_PastaCurvEffect, Newton_ea13_Survey, Sharma_Centelles+15, Schuetrumpf+15, Fattoyev+17, Pearson_ea20_Pasta, Oyamatsu_ea20, Xia_ea21_pasta, JiHuShen21, DinhThi+21a, DinhThi+21b, Pearson+22, Shchechilin_ea22_pasta, Parmar_ea22_pasta, Chamel_ea22_pasta,DinhThi_ea22_PastaProp,Pelicer_ea22}), while some papers suggested more complex structures on the base of numerical simulations \cite{Berry_ea16_Parking-garage, Schneider_ea14_Waffles,Newton_ea21_Glassy_pasta}.
In this paper we limit ourselves to cylindrical nuclear shapes (both inverse and normal).

The mantle affects neutron star evolution (e.g.,\ \cite{Caplan_Horowitz17_Colloq} for review), through its effect on the transport properties (e.g., \cite{Horowitz_Berry08_pastaTransp,Yakovlev15_PastaTransp,Schmitt_Shternin18_transpReview,Yakovlev_ea18_PastaBulkVisc,Nandi_Schramm18, Pelicer_ea23_PastaTransp}) and neutrino emissivity (e.g., \cite{Gusakov_ea04_PastaDUrca,Alloy_Menezes11,Schuetrumpf_ea20_PastaStructFact,Lin_ea20_NeutrinoPasta}).
However, in this paper we concentrate on another aspect of the mantle properties: the elasticity. Indeed, each pasta-like layer of the mantle is similar to liquid crystal and can support shear stresses  (see \cite{Pethick-Potekhin}, for example).  
Its elastic properties can affect present-days observations (e.g., quasi-periodic oscillations, observed after magnetar flares \cite{Sotani11_PastaConstr,Gearheart_ea11_PastaEffects,Sotani19_PastaTors} as well as constraints on quadrupole ellipticity supported by the crust \cite{Gearheart_ea11_PastaEffects}), and, as a part of the crust elasticity, can potentially affect observations of the third generation of gravitational wave detectors  (e.g.\ \cite{Biswas_ea19}, see, however, critique in \cite{Gittins_ea20_TidalElast}).

The seminal work on the elastic properties of the mantle was the paper \cite{Pethick-Potekhin}. Therein, the authors calculated elastic constants for cylindrical and planar nuclear shapes in the framework of compressible liquid drop model (CLDM).
Most of results were obtained analytically, using some simplified assumptions.
Elastic properties of the mantle were also studied by molecular dynamics \cite{Caplan_ea18_ElastPasta} and relativistic-mean-field model \cite{Xia_ea23_ElastPasta}; special attention were paid to the problem of effective shear modulus in a realistic case in which non-spherical nuclei are not well ordered, considering global hydrodynamical scales of a star, i.e.\ the mantle has `polycrystalline' structure \cite{Pethick_ea20_Pasta}.

In this paper we elaborate CLDM for elasticity of the mantle.
Namely, we use our recent work \cite{ZC23_shear_modulus}, where we have demonstrated that shear deformation of neutron star crust induces quadrupole deformation for the initially spherical nuclei and this effect reduces the effective shear modulus.
Here we generalize the approach of \cite{ZC23_shear_modulus} for cylindrical pasta phases and consider transversal shear, i.e.\ shear deformations in the plane, perpendicular to the symmetry axis of the cylinders. 
As for spherical nuclei, we demonstrate that the respective elastic constant (coefficient $C$ in notations of \cite{Pethick-Potekhin}) can be analytically calculated within simplified Wigner-Seitz approximation and the result agrees well with accurate lattice-based calculations \cite{Pethick-Potekhin}, if the same assumption (enforced circular shape of the cluster cross-section) is applied.
As pointed in \cite{Pethick-Potekhin}, shape of nuclear clusters can be affected by transversal shear. We take into account this effect analytically, allowing nuclear clusters to adjust their shape to minimize energy in response to the shear of the lattice.
We demonstrate that shear deformation indeed enforces nuclei clusters to have elliptical cross-section and, if one accounts for this effect, one gets $\sim 25\%$ reduction of the respective elastic constant at the most relevant parameters.
In Section 2, we present our formalism and derive the expression for transverse shear modulus for cylindrical clusters. In Section 3, we show our results and compare them with previous works. 
In Section 4 we present a brief summary of the results.

\section{Formalizm}

Nuclear clusters in the mantle have microscopical structure similar to that of liquid crystals resembling an intermediate state between solids and liquids in terms of elastic properties \cite{Pethick-Potekhin, Caplan_ea18_ElastPasta,Pethick_ea20_Pasta}. In what follows, we use terminology and notation of Ref.\ \cite{Pethick-Potekhin}. 
For the considered cylindrical phase, three elastic constants $B$, $C$, and $K_3$ are generally required to describe shear deformations (the compression is described by the forth constant, the bulk modulus, which can be easily calculated from the equation of state). The constant $B$ characterises the response to elongation along the cylindrical axis accompanied by compression in radial direction,  $K_3$ is associated with bending, while $C$ describes transverse shear. In this paper we limit ourselves to the elastic constant $C$.

\begin{figure}[h!]
	\begin{center}
	\includegraphics[width=6 cm]{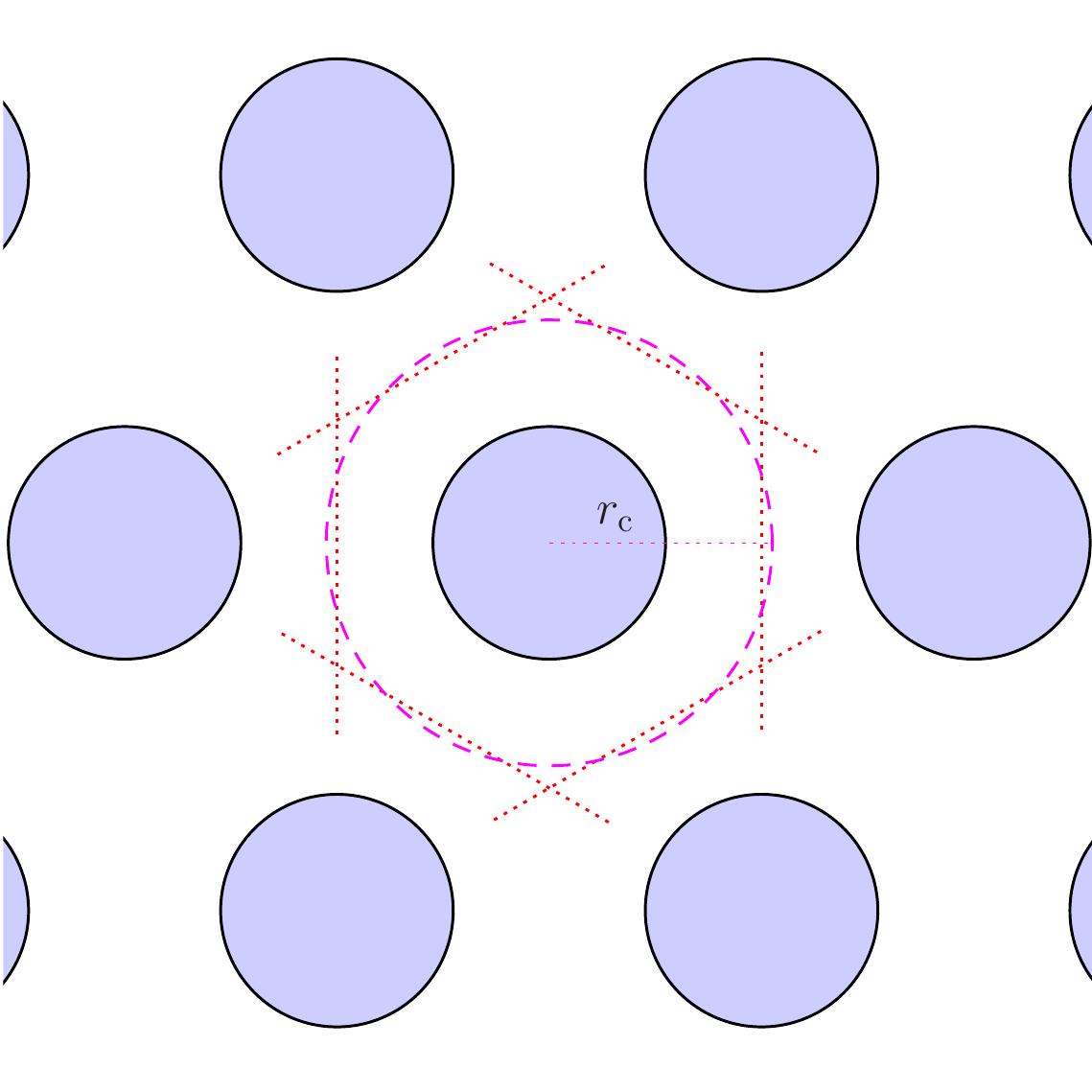}
	\ \ \includegraphics[width=6 cm]{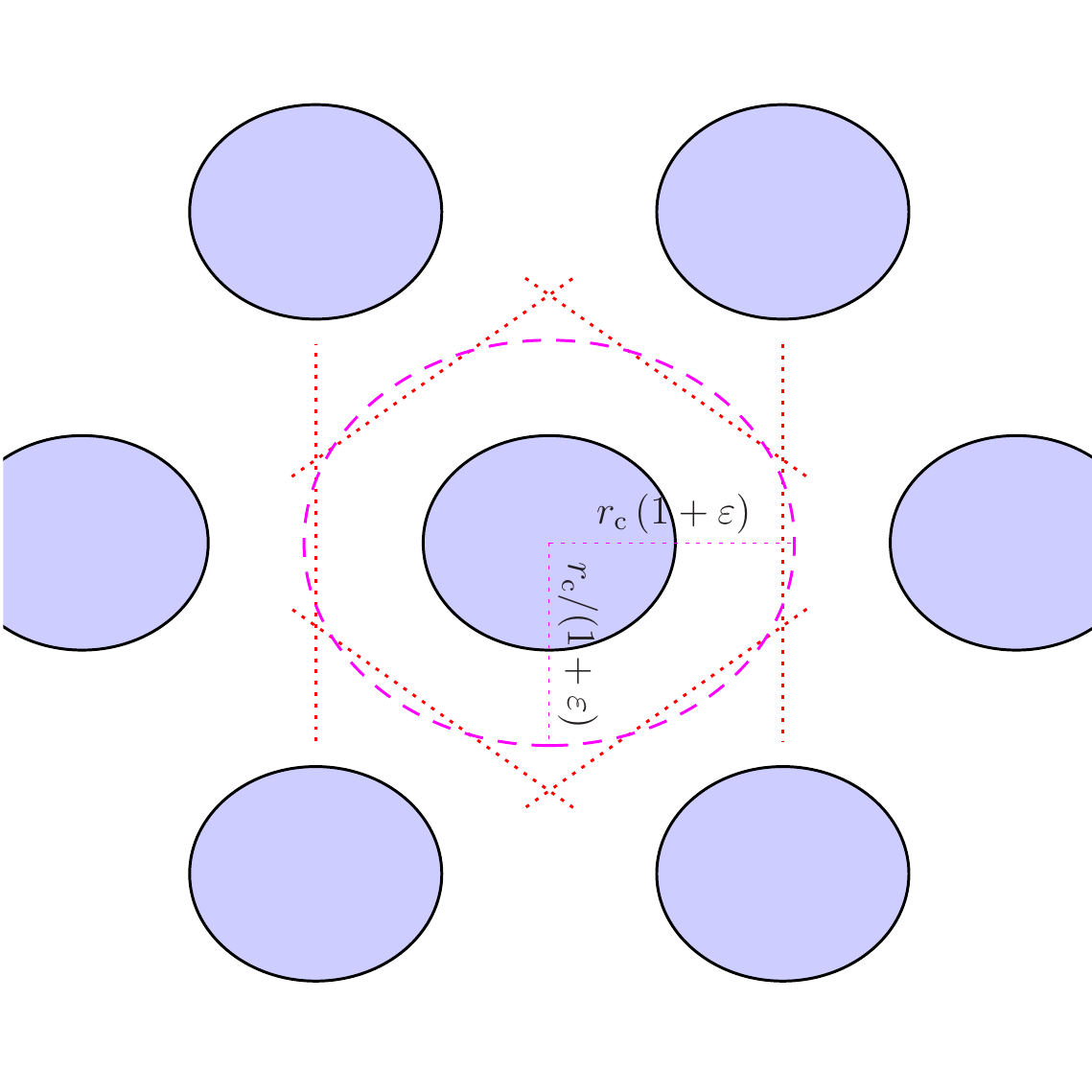}
	\end{center}
	\caption{Transversal cross-section of the lattice of cylindrical clusters (filled regions) in the spaghetti phase before (left panel) and after (right panel) transversal shear.
	Dotted lines represent accurate hexagonal boundaries of the  Wigner–Seitz cell, dashed lines are  circular/elliptical approximations of the cell.
	Change of the cluster cross-section in response to the applied shear is schematically shown at the right panel.
	}
\label{fig_Deform}
\end{figure}

To calculate the elastic constant $C$, we proceed in the same way as in \cite{ZC23_shear_modulus}.
First, we note that straightforward consideration of deformation for a system of cylindrical clusters, ordered into perfect static two-dimensional hexagonal lattice (as it was done in \cite{Pethick-Potekhin}), is equivalent to  deformation of one unit cell with periodic boundary conditions. It is obvious because the whole crystal (for both undeformed and deformed states) can be presented as a set of periodic copies of the unit cell.
Second, we replace a precise unit cell (regular hexagon) with a smoother figure: cylinder with a circular cross-section with radius  $r_\mathrm{c}$ and apply shear deformation 
\begin{eqnarray}
 x&\rightarrow& x\,(1+\varepsilon) \label{Deform_x}\\
 y&\rightarrow& y/(1+\varepsilon) \label{Deform_y}\\
 z&\rightarrow& z \label{Deform_z}
\end{eqnarray}
for this cell. 
Here $z$ axis is along the cylindrical axis and $\varepsilon$ is an infinitesimally small strain parameter. 
As a result of this volume-conserving deformation, the approximate unit cell for deformed state becomes a cylinder having an elliptical cross-section with semi-axes $r_\mathrm{c}(1+\varepsilon)$ and $r_\mathrm{c}/(1+\varepsilon)$  (see Figure \ref{fig_Deform}). High accuracy of the approach based on approximate unit cell is confirmed below by comparison with the results of \cite{Pethick-Potekhin} (for this aim we provide additional estimate of $C$, with  enforced circular shape of the cluster cross-sections, as it was assumed in \cite{Pethick-Potekhin}).

By definition (see equation (11) in \cite{Pethick-Potekhin}), for the specified deformation (equations \ref{Deform_x}--\ref{Deform_z}), the energy density change  is coupled with the elastic constant $C$ by
\begin{equation}
\delta w =2\,C \varepsilon^2. \label{C_def}
\end{equation}
Thus, it is enough to calculate the energy density change $\delta w$ to estimate $C$.

To calculate the energy density in deformed and undeformed states we apply compressible liquid drop model (CLDM) in the same way as in \cite{GusakovChugunov_Accr_Crust}, i.e.\ we parametrize surface parameters for plain interface  by neutron chemical potential $\mu_\mathrm{n}$, taking into account neutron adsorption (adsorption of protons vanishes due to the choice of the reference frame, see supplementary material in \cite{GusakovChugunov_Accr_Crust} for details), but neglect curvature corrections.
Below we provide derivations for non-inverted cylindrical phase,%
\footnote{Derivations for the inverted phase can be performed straightforwardly by substituting $u \rightarrow 1-u$, and treating $r_\mathrm{p}$ as the linear radius of the neutron phase, located in the center. \label{foot_invert}}
i.e., we consider a cell with neutron-proton cluster (with number densities $n_\mathrm{ni}$ and $n_\mathrm{pi}$) located in a central part, surrounded by neutron matter with number density $n_\mathrm{no}$ (see figure \ref{fig_Deform}).
In comparison with \cite{Pethick-Potekhin}, we allow the cluster cross-section in the $x,y$ plane to response to the applied shear. As long as we consider infinitesimal deformations, the cross-section in deformed state can be 
treated as ellipsoidal with semi-axes 
$r_\mathrm{p}(1+\varepsilon_\mathrm{p})$ and $r_\mathrm{p}/(1+\varepsilon_\mathrm{p})$.
Here $\varepsilon_\mathrm{p}$ is an infinitesimal parameter, which allows us to adjust cluster shape to minimize energy density. Below we consider the case when semi-axis $r_\mathrm{p}(1+\varepsilon_\mathrm{p})$ of cluster cross-section lies along $x$. We
have checked that it is this configuration, which corresponds to the minimal energy density within more general treatment, which allows the cluster deformation to be oriented arbitrary.
Note that, the volume fraction, occupied by the cluster  depends neither on $\varepsilon$ nor on $\varepsilon_\mathrm{p}$ and can be written as  $u\equiv (r_\mathrm{p}/r_\mathrm{c})^2$.
As usual, we assume the cell to be quasi-neutral, filled by degenerate electrons with uniform density in the whole cell. As a result, the energy density  is
\begin{equation}
w=u w_\mathrm{b}(n_\mathrm{ni},n_\mathrm{pi})
        +(1-u)w_\mathrm{b}(n_\mathrm{no},n_\mathrm{po}=0)
        +w_\mathrm{S}+w_\mathrm{C}+w_\mathrm e(n_\mathrm e).
\end{equation}
Here, $w_\mathrm{b}(n_\mathrm{n},n_\mathrm{p})$ is the bulk energy density of uniform nuclear matter at neutron and  proton number densities $n_\mathrm{n}$ and $n_\mathrm{p}$, 
$w_\mathrm e(n_\mathrm e)$ is the energy density of degenerate electrons with the number density $n_\mathrm e=u\,n_\mathrm{pi}$. 
Finally,  
$w_\mathrm{S}$ is energy density associated with surface energy of the cluster and $w_\mathrm{C}$  is the Coulomb energy density (averaged over cell).
These quantities are derived below.

As usual, within CLDM models, we assume that the internal parameters ($n_\mathrm{ni}$, $n_\mathrm{pi}$, $n_\mathrm{no}$,  $r_\mathrm{c}$, $r_\mathrm{p}$, $\varepsilon_\mathrm{p}$) are adjusted to minimize the energy density at given external parameters. In the case of deformed lattice, the external parameters are the baryon number density $n_\mathrm{b}$ and the shear parameter $\varepsilon$.

To derive the energy density associated with surface energy $w_\mathrm{S}$, we remind that the surface energy 
per unit area is $\sigma +\mu_\mathrm{n} \nu_\mathrm{S}$,
where $\sigma$ is the surface tension, $\nu_\mathrm{s}$ is the surface number density of the adsorbed neutrons (see \cite{lpr85,Haen_Potek_Yakov_book}, for example). The area of the cylindrical cluster is given by the product of the cluster height $2H$ to the perimeter (circumference) of the cluster cross-section $P$. It allows to write down the energy density, associated with surface energy, in the form
\begin{equation}
w_\mathrm{S}=\frac{(\sigma +\mu_\mathrm{n} \nu_\mathrm{S}) P}{\pi r^2_\mathrm{c}}.
\end{equation}

As written above, we assume cluster cross-section to be an ellipse with semi-axes 
$r_\mathrm{p}(1+\varepsilon_\mathrm{p})$ and $r_\mathrm{p}/(1+\varepsilon_\mathrm{p})$. Assuming $\varepsilon_\mathrm{p}>0$, the major semi-axis is $r_\mathrm{p}(1+\varepsilon_\mathrm{p})$ and
\begin{equation}
	P=4r_\mathrm{p}(1+\varepsilon_\mathrm{p})\int_0^{\pi/2}\sqrt{1-k^2\cos^2 t}\,\mathrm d t=4r_\mathrm{p}(1+\varepsilon_\mathrm{p})E(k^2),
\end{equation}
where $E(k)$ is the complete elliptic integral of the second kind (see, for example, \cite{Abramowitz_Stegun}) and $k=\sqrt{1-1/(1+\varepsilon_\mathrm p)^{4}}$ is eccentricity of the cluster cross-section. 
Using the  series for $E(k^2)$ (see, for example, equation (17.3.12) in \cite{Abramowitz_Stegun}) or straightforwardly applying equation (45) from \cite{Ramanujan1914} we obtain $P=2\pi\,r_\mathrm{p}\left(1+3\varepsilon_\mathrm{p}^2/4+O(\varepsilon_\mathrm{p}^3)\right)$,%
\footnote{Note, that terms up to $k^4$ should be included to obtain this expression.} leading to the final expression for the surface energy density
\begin{equation}
w_\mathrm{S}=\frac{2\,u\,(\sigma +\mu_\mathrm{n} \nu_\mathrm{S})}{r_\mathrm{p}}
\left(1+\frac{3}{4}\varepsilon_\mathrm{p}^2 \right).
\end{equation}
It is easy to check that the same equation also holds true for $\varepsilon_\mathrm{p}<0$ (for this aim it is enough to repeat derivations in this passage, taking into account that the major semi-axis for  $\varepsilon_\mathrm{p}<0$ is $r_\mathrm{p}/(1+\varepsilon_\mathrm{p})$ and thus eccentricity $k=\sqrt{1-(1+\varepsilon_\mathrm p)^{4}}$).

Within CLDM, calculation of the Coulomb energy density $w_\mathrm{C}$ reduces to the essentially electrostatic problem: calculation of the energy of neutral system composed of positively uniformly charged cylinder (protons inside the cluster) inserted into a negatively charged cylinder (electrons). The Coulomb energy density can be written in the form
\begin{equation}
w_\mathrm{C}=\frac{1}{2\pi r_\mathrm{c}^2}
\int \rho_\mathrm{p} (\bm{r})\varphi_\mathrm{p}(\bm r)  d^2 \bm {r} 
+\frac{1}{2\pi r_\mathrm{c}^2}
\int \rho_\mathrm{e} (\bm{r})\varphi_\mathrm{e}(\bm r)  d^2 \bm {r}
+\frac{1}{\pi r_\mathrm{c}^2}
\int \rho_\mathrm{p} (\bm{r})\varphi_\mathrm{e}(\bm r)  d^2 \bm {r},\label{e_C_gen}
\end{equation}
where the integrals are taken over vector $\bm r$ in the $x,y$ plane, $\varphi_\mathrm{p}(\bm r)$ and $\varphi_\mathrm{e}(\bm r)$ are electrostatic potentials, created by protons and electrons respectively;  $\rho_\mathrm{p} (\bm{r})$ and $\rho_\mathrm{e} (\bm{r})$ are charge densities of protons and electrons respectively.
While considering this electrostatic problem we assume $\rho_\mathrm{p} (\bm{r})=e\,n_\mathrm{pi}$ inside the internal cylinder (cluster) and  outside the internal cylinder $\rho_\mathrm{p} (\bm{r})=0$; similarly, $\rho_\mathrm{e} (\bm{r})=-e\,n_\mathrm{e}$  inside the external cylinder (cell) and $\rho_\mathrm{e} (\bm{r})=0$ outside the external cylinder.
Here  $e$ is elementary charge.

To apply (\ref{e_C_gen}), it is sufficient to know electrostatic potential inside a uniformly charged (charge density $\rho$) cylinder with elliptical cross-section (semi-axes $a_x$ and $a_y$ along $x$ and $y$ respectively). This potential can be written as (e.g.\ \cite{Kondratyev07_book}):
\begin{equation}
\phi=\pi\rho\left\{2\,a_x\, a_y \left[\ln\left(\frac{4H}{a_x+a_y}\right)+\frac{1}{2}\right]-\frac{2 \left(a_y\, x^2+a_x\,y^2\right)}{a_x+a_y}\right\}.
\end{equation}
Here $2H$ is a height of the cylinder, that is assumed to be large ($H\gg a_x$ and $H\gg a_y$).

Taking into account that ellipticities of the cell and cluster are infinitesimally small, it is straightforward to write down 
\begin{equation}
w_\mathrm{C}
=\frac{\pi}{2} u\,e^2 n^2_\mathrm{pi} r_\mathrm{p}^2
\left[
u-1-\ln(u)
+(2\,u-1)\varepsilon_\mathrm{p}^2
-2u\varepsilon_\mathrm{p}\varepsilon
+\varepsilon^2
\right]+\ldots \label{ec}
\end{equation}
where terms of the third and higher order in cluster and cell ellipticities are omitted.
As it should be, $H$ is cancelled out in the final expression.

Now, it is easy to write down equations for internal variables by minimizing the energy density at fixed $n_\mathrm{b}$ and $\varepsilon$.
As in the case of spherical nuclei, considered in \cite{ZC23_shear_modulus}, the majority of internal variables ($n_\mathrm{ni}$, $n_\mathrm{pi}$, $n_\mathrm{no}$,  $r_\mathrm{c}$, $r_\mathrm{p}$) can be derived at $\varepsilon=0$ up to corrections of order of $\varepsilon^2$,%
\footnote{Thanks to minimization procedure over internal variables, corrections of order of $\varepsilon^2$ contribute to the energy at order of $\varepsilon^4$, and thus can be neglected}
and only  $\varepsilon_\mathrm{p}\propto \varepsilon$:
\begin{equation}
0 =\pi u\,e^2 n^2_\mathrm{pi} r_\mathrm{p}^2
 \left[(2\,u-1)\varepsilon_\mathrm{p}
 -u\varepsilon\right]
 +\frac{3\sigma u}{r_\mathrm{p}}\varepsilon_\mathrm{p}.
 \label{ep_minimize}
\end{equation}

Equations for $\varepsilon=0$ are well known (e.g., \cite{GusakovChugunov_Accr_Crust} in case of spherical nuclei);
their physical meaning are chemical equilibrium within the cell, beta-equilibrium,  pressure balance for the cluster, and the equation for  equilibrium size of the cell. The latter is often called  virial theorem (see, for example, \cite{Ravenhall_pasta,Hashimoto_ea84}), and it is the only equation, which would be required below in explicit form:
\begin{equation}
\frac{\sigma u}{r_\mathrm{p}} =\frac{\pi}{2} \rho^2_\mathrm{p}r^2_\mathrm{p}u \left(u-1-\ln(u) \right).
\end{equation}

Virial theorem allows to exclude $\sigma$, removing  explicit dependence of the subsequent results on the nuclear physical model.
Indeed, the bulk contributions to the total variation of energy are cancelled out and it can be written in the form
\begin{equation}
\delta w= \pi \rho^2_\mathrm{p}r^2_\mathrm{p}u \left[\frac{1}{2}\varepsilon^2-u\varepsilon \varepsilon_\mathrm{p}+
\frac{7u-5-3\,\ln(u)}{4}\,\varepsilon_\mathrm{p}^2
\right].
\label{de}
\end{equation}
In the case of $\varepsilon=0$, considered in \cite{Iida_ea01}, it agrees with the sum of equations (6)-(9) from that work.
For $\varepsilon\ne0$ the optimal cluster deformation parameter, given by solution of Eq.\ (\ref{ep_minimize}), is
\begin{equation}
\varepsilon_\mathrm{p}=\frac{2u}{7u-5-3\,\ln(u)} \varepsilon. \label{ep}
\end{equation}

Substituting (\ref{ep}) into (\ref{de}) and using (\ref{C_def})  we come to the final expression for the elastic constant for the transverse shear:
\begin{equation}
C=\frac{\pi \rho_\mathrm{p}^2r^2_\mathrm{p}u}{4} \left(1-\frac{2u^2}{7u-5-3\,\ln(u)} \right). \label{C}
\end{equation}

It is worth to stress, that the $\partial^2 \delta w/\partial \varepsilon_\mathrm{p}^2$, calculated at the optimal value of $\varepsilon_\mathrm{p}$, is positive for arbitrary filling factor.
Thus the clusters can indeed adjust their shape and become stable with respect to additional infinitesimal deformations.
Stability of cylinders in undeformed lattice ($\varepsilon=0$) was proven in \cite{Iida_ea01}.
A similar result was obtained in Ref.\ \cite{Zemlyakov_ea22_stab}, according to which the spherical nuclei in the crust are stable with respect to infinitesimal quadrupole deformations, and thus transition from the crust to the mantle is not associated with absolute instability of spherical nuclei.

To compare our approach with the results of Ref.\ \cite{Pethick-Potekhin}, we also consider approximation of enforced circular shape of cluster cross-section ($\varepsilon_\mathrm{p}=0$, not adjusted to the deformation).
It leads to a simplified estimate of transverse shear modulus
\begin{equation}
C^\mathrm{sp}=\frac{\pi \rho_\mathrm{p}^2r^2_\mathrm{p} u}{4}. \label{Csp}
\end{equation}

\section{Results}\label{Sec_NumRes}

\begin{figure}[H]
	\includegraphics[width=0.95\textwidth]{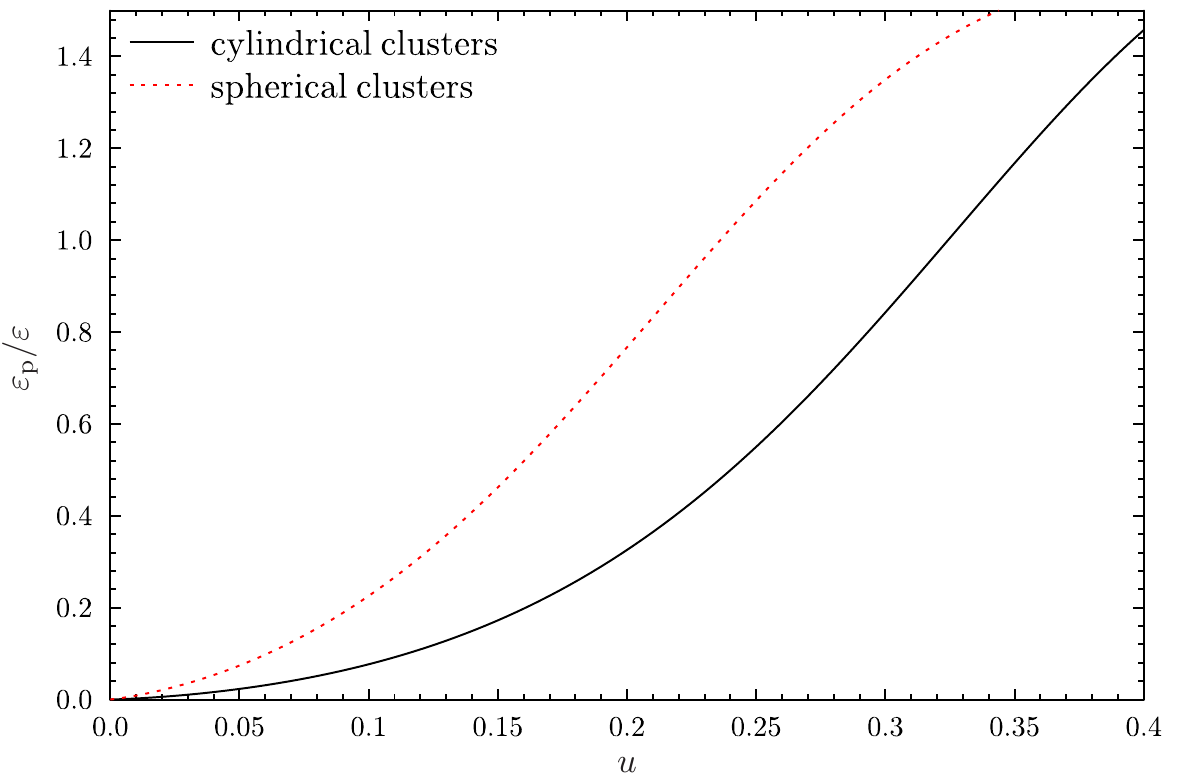}
	\caption{The ratio of cluster deformation parameter $\varepsilon_\mathrm p$ to the shear parameter $\varepsilon$ as a function of the filling factor $u$.
		Solid line is for cylindrical clusters (this work), dotted line is for spherical clusters (Ref.\ \cite{ZC23_shear_modulus})
		\label{fig_epsp}
	}
\end{figure}
In figure \ref{fig_epsp} we demonstrate the ratio $\varepsilon_\mathrm p/\varepsilon$, given by Eq.\ (\ref{ep}), as a function of filling factor $u$ (solid line). For a typical filling factor of the cylindrical phases, $u\approx 0.2\div0.3$ (e.g., \cite{Hashimoto_ea84,Oyamatsu_ea20}), the cluster deformation evolves from $\varepsilon_\mathrm p\approx 0.3\varepsilon$  to the value, which is close to $\varepsilon$ ($\varepsilon_\mathrm p\approx 0.85\varepsilon$). For comparison, the dotted line indicates similar ratio for spherical clusters, which was obtained in Ref.\ \cite{ZC23_shear_modulus}. One can see that this ratio depends on the shape of the clusters.
In particular, for $u\approx 0.2$, which is typical for transition from spherical to cylindrical clusters (e.g., \cite{Oyamatsu_ea20}), the shape of spherical clusters is approximately two times more sensitive to the shear deformations, than the shape of cylindrical clusters.

\begin{figure}[H]
	\includegraphics[width=0.95\textwidth]{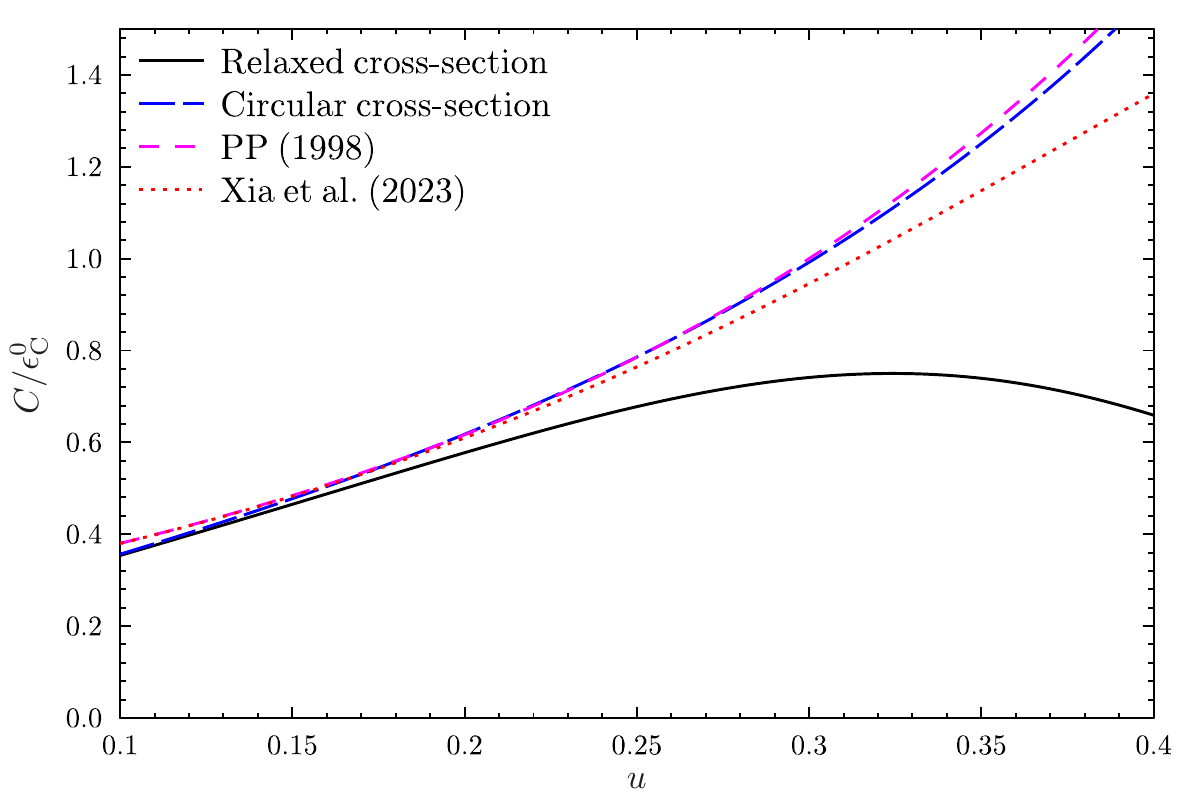}
	\caption{Transversal shear modulus $C$ for 'spaghetti' phase of neutron star mantle as a function of filling factor. 
		The solid and long-dashed lines are our results, with optimized  (Eq.\ \ref{C}) and circular  (Eq.\ \ref{Csp}) shapes of cluster cross-section respectively.
 		Short dashes represent the fit from \cite{Pethick-Potekhin},
		while the dotted line corresponds to the fit from \cite{Xia_ea23_ElastPasta}.
		\label{fig_compare}
	}
\end{figure}

To compare our results with previous works, in figure \ref{fig_compare} we demonstrate the transverse shear modulus, normalized to the Coulomb energy density in the non-deformed state $w^0_\mathrm{C}$ (see Eq.\ (\ref{ec}) for $\varepsilon=\varepsilon_\mathrm{p}=0$) as a function of filling factor $u$.
The solid line represents our final expression (\ref{C}), while the long-dashed line is for Eq.\ (\ref{Csp}) ignoring the change of the cross-section shape.
The latter agrees well with the fit provided in  \cite{Pethick-Potekhin} (see short dashes), which was based on calculations with accurate treatment of hexagonal (honeycomb) lattice, but with fixed circular cross-section of the cylindrical clusters. Note, that the authors of \cite{Pethick-Potekhin} fitted their numerical results for reasonable filling factor $u\sim 0.2\div0.3$ (e.g., \cite{Hashimoto_ea84,Oyamatsu_ea20}).

Recently the authors of Ref. \cite{Xia_ea23_ElastPasta} analysed elastic properties of nuclear pasta in a full three-dimensional geometry.
Their calculations were performed within  Thomas-Fermi approach based on relativistic-mean-field model, and thus should include change of the cluster cross-section shape with increasing deformation.  The fit for the transversal shear modulus suggested in \cite{Xia_ea23_ElastPasta} is shown by dotted line in Fig.\ \ref{fig_compare}.
Obviously, it predicts much smaller suppression of $C$, than our result (\ref{C}).

However, we should note that rather large deformations (from a hexagonal to a simple lattice)
were applied to extract $C$  in \cite{Xia_ea23_ElastPasta}. 
Namely, the authors of  \cite{Xia_ea23_ElastPasta} performed numerical simulations for a wide set of  deformation parameters at fixed baryon number density and fit the results by their equation (13).
As a result, their equation (13) reproduces well the deformation energy for the whole deformation path from a hexagonal to another hexagonal lattice via simple lattice (see figure 3 in \cite{Xia_ea23_ElastPasta}). But it has only one fitted parameter (elastic constant, which we will be denoted as $C_\mathrm{13}$). It seems to be mostly affected by the energy difference between hexagonal and simple lattices (which is indeed perfectly reproduced, according to figure 3 in \cite{Xia_ea23_ElastPasta}), but not to the details of infinitesimal deformations of hexagonal lattice. 

Let us note, that there should be no elliptical deformation of the cluster cross-sections for both simple and hexagonal lattices due to symmetry. Thus the energy difference between these lattices can be reasonably estimated assuming circular cluster shape.
As a result, the effects of elliptical deformation of cluster cross-section can be suppressed in $C_\mathrm{13}$, overestimating the elastic constant $C$.
Indeed, our attempt to directly fit numerical data, shown in figure 3 of \cite{Xia_ea23_ElastPasta} for small transversal shear, leads to $C\sim 0.8C_\mathrm{13}$, i.e.  20\% smaller than the estimate of the shear modulus suggested in \cite{Xia_ea23_ElastPasta} (the red line at the same plot). Thus, we expect that the actual transversal shear modulus for infinitesimal deformation of hexagonal lattice can be smaller than that given by the final fit presented in \cite{Xia_ea23_ElastPasta} being closer to our Eq.\ (\ref{C}).

Unfortunately, we cannot check this hypothesis directly  because the filling factor for the baryon number density $n_\mathrm{b}=0.06$~fm$^{-3}$, analyzed in their figure 3, is not quoted in \cite{Xia_ea23_ElastPasta}.
However, we can estimate the filling factor using figures in Ref. \cite{Oyamatsu_ea20}.
Indeed, figure 3 in \cite{Xia_ea23_ElastPasta} is plotted for the slope of the symmetry energy $L = 41.34$~MeV. Thus, according to figure 4 of \cite{Oyamatsu_ea20}, $n_\mathrm{b}=0.06$~fm$^{-3}$ should be close to the transition from spherical to cylindrical clusters, and according to figure 3 of same work it should correspond to $u\approx 0.2$.
In this case, our Eq.\ (\ref{C}) predicts $C$ to be suppressed by $\sim 6.5\%$ with respect to results for fixed (circular) cluster cross-section, which almost coincides with $C_{13}$ for $u\approx 0.2$.
Formally, it disagrees with the estimate $C\sim 0.8C_\mathrm{13}$ mentioned above, and suggests that our Eq.\ (\ref{C}), in fact, overestimates $C$. However, due to qualitative nature of estimates in this passage we prefer to treat this conclusion as very tentative.

\begin{figure}
	\includegraphics[width=0.95\textwidth]{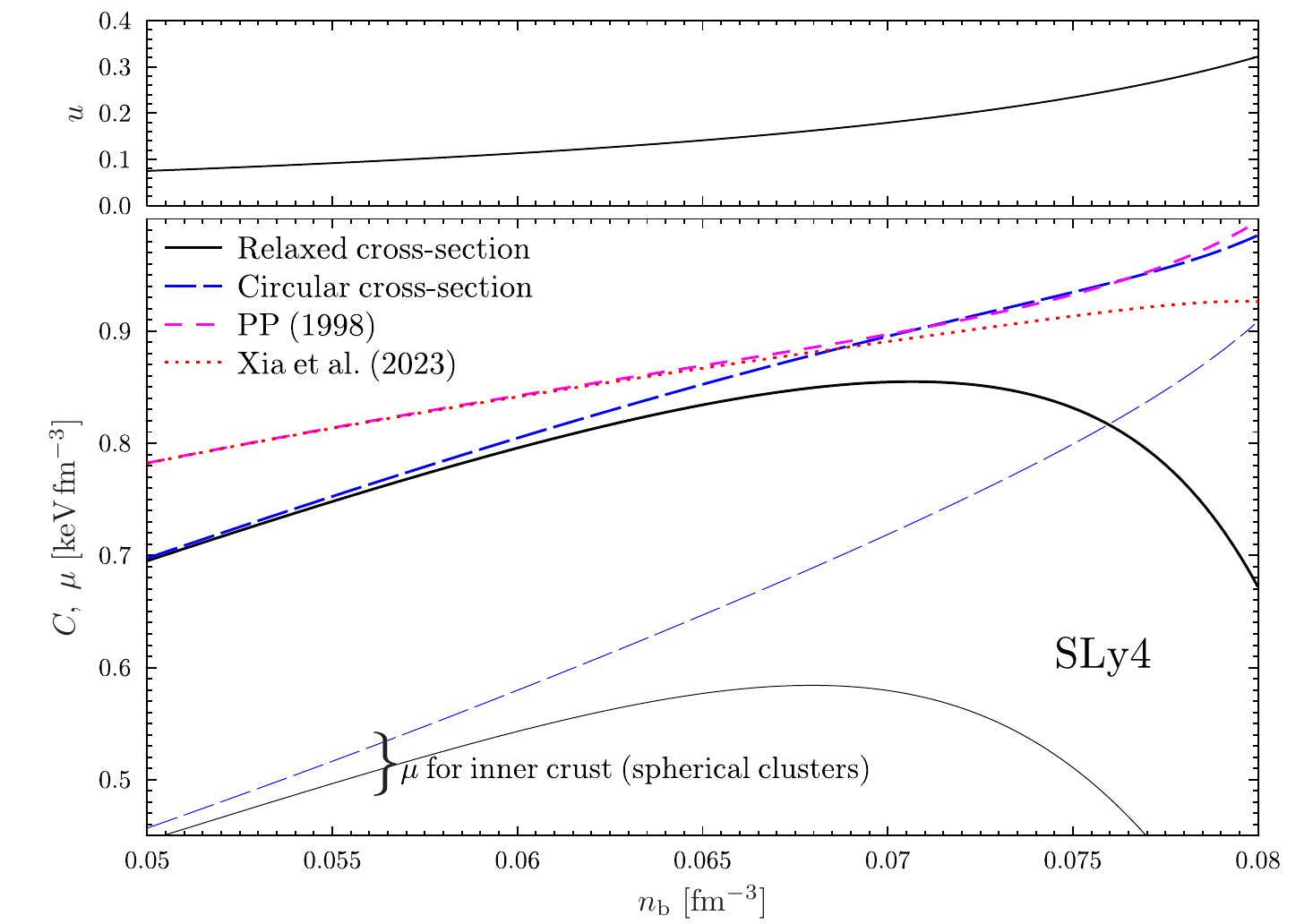}
	\caption{
				Transversal shear modulus $C$ for 'spaghetti' phase of neutron star mantle in physical units as a function of baryon number density $n_\mathrm{b}$.
	As in figure \ref{fig_compare}, the solid and long-dashed lines are our results, calculated with optimized  (Eq.\ \ref{C}) and circular  (Eq.\ \ref{Csp}) shapes of cluster cross-section respectively.
	Short dashes represent the fit \cite{Pethick-Potekhin};
	the dotted line corresponds to the fit  \cite{Xia_ea23_ElastPasta}.
	In addition, thin lines represent effective shear modulus for spherical nuclear clusters, as it was calculated in \cite{ZC23_shear_modulus}. Namely, thin solid line accounts for nuclei deformation (equation 9 in \cite{ZC23_shear_modulus}), while thin dashed line neglects this effect (equation 6 in \cite{ZC23_shear_modulus}).
	The upper panel represents cluster volume fraction $u$.
	SLy4 nucleon interaction model \cite{Chabanat_ea97_SLY4} was used to calculate bulk energy contribution; the surface tension were calculated according to fit of the fourth-order ETF calculations for the SLy4 model, provided by N.N.\ Shchechilin (the same fit was used in \cite{Shchechilin_ea22_pasta}).
		\label{fig_phys}
	}
\end{figure}

As discussed in \cite{Shchechilin_ea22_pasta}, the actual shape of nuclear clusters at given number density is rather model dependent, even if nucleon interaction model is fixed. For example, for Skyrme Lyon nucleon interaction model  SLy4 \cite{Chabanat_ea97_SLY4}, which was analyzed in \cite{Shchechilin_ea22_pasta}, the cylindrical clusters are the most energetically favourable at baryon number density $0.056\mbox{~fm}^{-3}\lesssim n_\mathrm{b}\lesssim 0.075\mbox{~fm}^{-3}$, if energy density is calculated within ETF approach, while CLDM predicts it to be favourable only in a narrow region $0.074\mbox{~fm}^{-3}\lesssim n_\mathrm{b}\lesssim 0.076\mbox{~fm}^{-3}$. Thus, to illustrate our results in physical units (see figure \ref{fig_phys}) we use rather wide $n_\mathrm{b}$ region; following \cite{Shchechilin_ea22_pasta}, we applied
SLy4 nucleon interaction model  for numerical estimates.
As in figure \ref{fig_compare}, the solid line represents our final expression (\ref{C}), while the long-dashed line is for equation (\ref{Csp}), which ignores the change of the cross-section shape.
Short dashes correspond to the fit provided in  \cite{Pethick-Potekhin}, while dotted line is for fit \cite{Xia_ea23_ElastPasta}. Note, that both fits present the ratio of $C$ to the Coulomb energy density and in our calculations we apply CLDM \cite{GusakovChugunov_Accr_Crust} to calculate the latter.
The upper panel demonstrates cluster volume fraction $u$. As discussed in \cite{Haen_Potek_Yakov_book}, the latter quantity can be estimated in so-called bulk approximation, being weakly dependent on the shape of the clusters and the applied approach (ETF or CLDM).

For the whole indicated density region our results predict transversal shear modulus $C$ to be lower, than according to fits from Refs.\ \cite{Pethick-Potekhin} and \cite{Xia_ea23_ElastPasta}. For low baryon number density $n_\mathrm{b}\lesssim 0.065$~fm$^{-3}$ it is associated with low  volume fraction  $u\lesssim 0.15$, where 
fit \cite{Xia_ea23_ElastPasta} almost coincides with \cite{Pethick-Potekhin}. However, the latter overestimates original numerical calculations of that work for $u\lesssim 0.15$ (see figure 2 in \cite{Pethick-Potekhin}).
Thus, we expect that our equation (\ref{C}) is, in fact, more accurate than fits of Refs.\ \cite{Pethick-Potekhin} and \cite{Xia_ea23_ElastPasta} at least for low density region.
The larger the density, the larger is effect of relaxation of the cluster cross-section. For $n_\mathrm{b}\gtrsim 0.07$~fm$^{-3}$ it leads to decrease of $C$ with increase of $n_\mathrm{b}$, in spite of the fact that $C^\mathrm{sp}$, calculated neglecting this effect, increases with increase of  $n_\mathrm{b}$.

For comparison,  in figure \ref{fig_phys} we also plotted effective shear modulus $\mu$ for inner crust matter, assuming spherical shape of the clusters \cite{ZC23_shear_modulus}.  Thin solid line was calculated taking into account relaxation of the cluster shape,  while dashed line neglects this effect.
One can see that effective shear modulus $\mu$ is lower than the transversal shear modulus $C$ for the same baryon number density. In principle, it can lead to a jump of elastic properties at the inner crust/mantle boundary. However, one should take in mind, that it is rather tricky to estimate effective shear modulus of ``polycrystalline'' pasta phases (see \cite{Pethick_ea20_Pasta} for discussion). In particular, it can vanish,  if pasta elements are spatially uniform (it is because the elastic constant, responsible for deformation along the pasta structure, is zero in a strictly uniform case). So we refrain from speculation on possible observational consequences of the difference between $C$ and $\mu$.

\section{Summary and discussion}

We analytically derived transversal shear modulus $C$ for the sphaghetti phase of nuclear pasta
(see Eq.\ \ref{C}, result for the inverted phase can be obtained using footnote \ref{foot_invert}).
Following to the predictions of Ref. \cite{Pethick-Potekhin}, we demonstrated that shear deformation affects the shape of nuclear clusters (the transversal cluster cross-section becomes elliptical) and we took this effect into account.
It is worth to note, that our final expression for $C$ is explicitly independent of the parameters of inter-nucleon interaction.

Our results agree with the well-known fit \cite{Pethick-Potekhin}, if one neglects cluster deformation, as it was done in \cite{Pethick-Potekhin}.
Accounting for cluster deformation reduces $C$ (for example, $C$ is reduced by 25\% at the filling factor $u\sim 0.3$).
However, the practical fit, suggested in the recent paper \cite{Xia_ea23_ElastPasta} on the base of numerical simulations in fully three-dimensional geometry, predicts much smaller reduction of $C$ than in Ref. \cite{Pethick-Potekhin}. As we argued in Section \ref{Sec_NumRes}, this difference can be associated with numerical procedure in \cite{Xia_ea23_ElastPasta}. Namely, the practical expression in \cite{Xia_ea23_ElastPasta}, in fact, suggests a parameter for their equation (13), which allows one to describe deformation energy for strong transversal deformations, but is not, however, very precise for smaller deformations.

Finally, we should remind that our approach is not exact by construction, being based on two simplifications: (1) we neglect curvature corrections to surface tension;
(2) we apply ellipsoidal approximation for the cell cross-section.
Both of these approximations can affect  $C$, and we plan to check their importance in subsequent studies.


\vspace{6pt} 



\authorcontributions{Derivations N.A.Z; formulation of the problem, supervision A.I.C. All authors have read and agreed with the
published version of the manuscript.
}

\funding{This research received no external funding.}

\dataavailability{Not applicable.} 

\acknowledgments{We thank Nikolai N. Shchechilin for providing us with the fit of the fourth-order ETF calculations of the surface tension for the SLy4 model.}

\conflictsofinterest{The authors declare no conflict of interest.} 



\abbreviations{Abbreviations}{
The following abbreviations are used in this manuscript:\\

\noindent 
\begin{tabular}{@{}ll}
CLDM & Compressible liquid drop model\\
ETF & Extended Thomas-Fermi
\end{tabular}
}


\begin{adjustwidth}{-\extralength}{0cm}

\reftitle{References}

\end{adjustwidth}
\end{document}